\documentclass[prd,aps,preprint,nofootinbib,superscriptaddress]{revtex4}
\usepackage{epsfig}
\usepackage{amsmath}
\input{epsf}

\begin{document}


\vspace*{2cm}
\title{QCD Evolution of the Transverse Momentum Dependent Correlations}

\author{Jian Zhou}
\affiliation{School of Physics, Shandong University, Jinan, Shandong 250100, China}
 \affiliation{ Nuclear Science Division, Lawrence Berkeley National Laboratory, Berkeley, CA 94720}

\author{Feng Yuan}
\affiliation{Nuclear Science Division, Lawrence Berkeley National Laboratory, Berkeley, CA 94720}
\affiliation{RIKEN BNL Research Center, Building 510A, Brookhaven National Laboratory, Upton, NY 11973}
\author{Zuo-Tang Liang}
\affiliation{School of Physics, Shandong University, Jinan, Shandong 250100, China}

\begin{abstract}
We study the QCD evolution for the twist-three quark-gluon
correlation functions associated with the transverse momentum odd
quark distributions. Different from that for the leading twist quark
distributions, these evolution equations involve more general
twist-three functions beyond the correlation functions themselves.
They provide important information on nucleon structure, and can be
studied in the semi-inclusive hadron production in deep inelastic
scattering and Drell-Yan lepton pair production in $pp$ scattering
process.
\end{abstract}

\maketitle

Semi-inclusive hadronic processes have attracted much theoretical
interest in recent years, where the so-called transverse momentum
dependent (TMD) parton distributions and fragmentation functions can
be studied
\cite{Brodsky:2002cx,Collins:2002kn,Ji:2002aa,Boer:2003cm,JiMaYu04,ColMet04,Mulders:1995dh,{Boer:1997nt}}.
These functions generalize the original Feynman parton picture,
where the partons only carry longitudinal momentum fraction of the
parenting (final state) hadron. They certainly provide further
important information on nucleon structure, and are crucial to
understand the novel spin phenomena, such as the single transverse
spin asymmetry
(SSA)~\cite{Brodsky:2002cx,Collins:2002kn,Ji:2002aa,Sivers:1990fh,{Anselmino:1994tv}}.

Important aspects of the TMD parton distributions have been explored
in the last few years, such as the gauge property and the crucial
role of the initial/final state interaction for the nonzero Sivers
quark distribution leading to the SSA in Semi-inclusive hadron
production in deep inelastic scattering (SIDIS) and Drell-Yan lepton
pair production processes~\cite{Brodsky:2002cx,Collins:2002kn,Ji:2002aa,Boer:2003cm}.
Further study has shown that this
mechanism is uniquely related to the twist-three quark-gluon
correlation approach for the SSA phenomena~
\cite{qiu,Efremov,Eguchi:2006qz,Eguchi:2006mc,new}. In particular,
these two approaches are unified to describe the same
physics in the overlap region where both apply~\cite{jqvy}.

At the leading order, there are eight independent TMD quark distributions, depending
on the polarizations of the nucleon and the quark~\cite{Mulders:1995dh,Boer:1997nt}.
Three of them are called $k_\perp$-even distributions. After integrating over transverse momentum,
they produce the leading-twist quark distributions, including the spin average,
longitudinal polarized, and transversity quark distributions~\cite{Jaffe:1991ra}. The rest five
distributions are called $k_\perp$-odd distributions. Upon integral over
the transverse momentum, they will vanish in the quark correlation matrix.
In this paper, we are interested in four of these $k_\perp$-odd TMD quark
distributions. They can be defined from the following matrix,
\begin{eqnarray}
    {\cal M}^{\alpha\beta} (x,k_\perp)&=&   P^+\int
        \frac{d\xi^-}{2\pi}e^{ix\xi^-P^+}\int
        \frac{d^2b_\perp}{(2\pi)^2} e^{-i\vec{b}_\perp\cdot
        \vec{k}_\perp} \, \left\langle
PS\left|\overline{\mit \Psi}_v^\beta(0)
        {\mit \Psi}_v^\alpha(\xi^-,0,\vec{b}_\perp)\right|PS\right\rangle\ ,
\end{eqnarray}
where $x$ is the longitudinal momentum fraction of the proton
carried by the quark and $k_\perp$ is the transverse momentum.
In the above definition we have chosen $P=(P^+,0^-,0_\perp)$ which
is along the momentum direction of the proton, $S$ is the polarization vector, and ${\mit
\Psi}_v(\xi)$ is defined as
${\mit \Psi}_v(\xi) \equiv {\cal L}_{v}(-\infty;\xi)\psi(\xi)\ , $
with ${\cal L}_{v}$ the gauge link. This gauge link contains the light-cone gauge link
contribution and the transverse gauge link contribution at the spatial
infinity~\cite{Ji:2002aa}. We have chosen it
goes to $-\infty$, indicating that we adopt the
definition for the TMD quark distributions for the Drell-Yan process
\cite{Brodsky:2002cx,Collins:2002kn,Ji:2002aa}.
The four $k_\perp$-odd TMD quark distributions can
be obtained by the following expansion of the above
quark correlation matrix~\cite{Mulders:1995dh,Boer:1997nt},
\begin{eqnarray}
{\cal M}&=&\frac{1}{2M}\left[
g_{1T}(x,k_\perp)\gamma_5\not\!
P(\vec{k_\perp}\cdot \vec{S}_\perp)+f_{1T}^\perp(x,k_\perp)\epsilon^{\mu\nu\alpha\beta}\gamma_\mu
P_\nu k_\alpha S_\beta
\right.\nonumber\\
&&\left.+h_{1L}(x,k_\perp)\lambda i\sigma_{\mu\nu}\gamma_5 P^\mu k_\perp^\nu+
h_1^\perp(x,k_\perp)\sigma^{\mu\nu}k_\mu P_\nu+ \cdots \right]\ , \label{tmdpar}
\end{eqnarray}
where $M$ is the nucleon mass and the interpretations of the four
$k_\perp$-odd TMD quark distributions are: $g_{1T}$ and
$f_{1T}^\perp$ represent a longitudinal polarized and unpolarized
quark distributions in a transversely polarized nucleon,
respectively; $h_{1L}$ and $h_{1}^\perp$ represent transversely
polarized quark distributions in a longitudinal polarized and
unpolarized nucleon target, respectively. They are $k_\perp$-odd
distributions, i.e., after integrating over the transverse momentum,
the above expansion will vanish. However, if we weight the integral
with transverse momentum, the above matrix will lead to a set of
quark-gluon correlation functions at the twist-three level. These
correlation functions can be calculated as transverse
momentum-moment of the above four $k_\perp$-odd TMD quark
distributions. The last $k_\perp$-odd TMD quark distribution
$h_{1T}^\perp$ represents a correlated transversely polarized quark
distribution in a transversely polarized nucleon target, and is
related to the twist-four quark-gluon correlation function. We will
not discuss this function in this paper.

As mentioned above, the transverse-momentum-moment of the above four TMD quark distributions
define the following transverse momentum dependent correlation functions in nucleon,
\begin{eqnarray}
\int d^2k_\perp \frac{\vec{k}_\perp^2}{2 \pi M^2} f_{1T}^\perp(x,k_\perp)&=&T_F(x) \ ,
~~~\int d^2k_\perp \frac{\vec{k}_\perp^2}{2  M^2} g_{1T}^\perp(x,k_\perp)=\tilde g(x) \ , \\
\int d^2k_\perp \frac{\vec{k}_\perp^2}{2 \pi M^2} h_{1}^\perp(x,k_\perp)&=&T_F^{(\sigma)}(x) \ ,
~~~\int d^2k_\perp \frac{\vec{k}_\perp^2}{2  M^2} h_{1L}^\perp(x,k_\perp)=\tilde h(x) \ .
\end{eqnarray}
We emphasize again that the above TMD quark distributions follow their
definitions in Drell-Yan process. If we choose those for the semi-inclusive DIS
process, the above two equations associated with $T_F(x)$ and $T_F^{(\sigma)}(x)$
shall change signs.
These correlation functions are related to more general quark-gluon correlation
functions. For example,  $T_F(x)$ and $T_F^{(\sigma)}(x)$ are diagonal parts
of the general quark-gluon correlation functions $T_F(x_1,x_2)$ and $T_F^{(\sigma)}(x_1,x_2)$
which are responsible for the single spin asymmetry in hadronic process~\cite{qiu,Zhou:2008fb}:
$T_F(x)\equiv T_F(x,x)$ and $T_F^{(\sigma)}(x)\equiv T_F^{(\sigma)}(x,x)$.\footnote{
For the convenience of our presentation, we have changed the normalizations
for $T_F(x_1,x_2)$ and $T_F^{(\sigma)}(x_1,x_2)$ by a factor
of $1/2\pi M$ as compared to those in~\cite{jqvy,{Zhou:2008fb}}.}
They can be defined through the following correlation matrix,
\begin{eqnarray}
M_{F \alpha\beta}^\mu(x)\equiv \int \frac{dy^-}{2 \pi}
\frac{dy_1^-}{2\pi} e^{ixP^+ y^-}  \langle PS |
\bar{\psi}_\beta(0)  g F^{+\mu}(y_1^-)\psi_\alpha(y^-) | PS
\rangle  \ ,
\end{eqnarray}
where $\mu $ is a transverse index, $ F_{+\mu}$ the gluon field tensor,
and the gauge link has been suppressed. Its decomposition
contains the contribution from $T_F(x)$ and $T_F^{(\sigma)}(x)$,
\begin{eqnarray}
M_{F \alpha\beta}^\mu(x)&=& \frac{M}{2} \left [
T_F(x)\epsilon^{\nu\mu}_\perp S_{\perp \nu} p\!\!\!/
+T_F^{(\sigma)}(x) i\gamma_\perp^\mu p\!\!\!/  \right ] \ .
\end{eqnarray}
Similarly, we can calculate the other two correlation functions by~\cite{Boer:2003cm},
\begin{eqnarray}
\tilde{M}_{F \alpha\beta}^\mu(x)&=& \frac{M}{2} \left [
\tilde{g}(x) S_{\perp}^{\mu} \gamma_5 p\!\!\!/
+\tilde{h}(x) \lambda \gamma_5 \gamma_\perp^\mu p\!\!\!/  \right ] \ .
\end{eqnarray}
where $\tilde{M}_{F \alpha\beta}^{\mu}$ is defined as,
\begin{eqnarray}
\tilde{M}_{F \alpha\beta}^{\mu}(x)&=&
\int\frac{d\xi^-}{2\pi}e^{i\xi^-xP^+}\langle
PS|\overline\psi_\beta(0)
\left \{ i{D_\perp}^\mu(\xi^-)-\int_{\xi^-}^{-\infty} d\zeta^-g F^{+\mu}(\zeta^-) \right \}
\psi_\alpha(\xi^-)|PS\rangle \ .
\end{eqnarray}
Applying the time-reversal invariance, we
find the above definition of $\tilde g$ is the same as that
in \cite{Eguchi:2006qz}, except a normalization factor 2.

The above four correlation functions are subsets of more general twist-three
quark-gluon correlation functions~\cite{Ellis:1982cd,Jaffe:1991ra}: $G_D(x,y)$,
$\tilde G_D(x,y)$, $H_D(x,y)$ and $E(x,y)$.
These twist-three functions and their contributions to the inclusive DIS and
Drell-Yan lepton pair productions
have been under intense investigations in the last two decades (see for example~\cite{Jaffe:1991ra}).
The above four correlation functions Eqs.~(3,4), however, will enter in the
transverse momentum weighted cross sections in the semi-inclusive
hadron production in DIS and Drell-Yan lepton pair production in $pp$
collisions~\cite{Mulders:1995dh,Boer:1997nt,{Bacchetta:2006tn}}.
They will provide additional information on the quark-gluon correlations in nucleon,
and will be complementary to those studied in the inclusive
DIS and Drell-Yan processes. Recent experimental developments
will help to pin down these contributions, and build strong physics
associated with these correlation functions~\cite{sidis}.

One of the important questions remained to be answered is
the scale evolution for these correlation functions. The evolution
equation controls the energy dependence of the associated
observables~\cite{ap}. For example, with the evolution equations, we will be
able to compare the single spin asymmetries coming from the same
quark-gluon correlation function $T_F(x)$ in hadronic processes
at different energy experiments.
General structure of the evolution equations for the twist-three
quark-gluon correlation functions has been known in the literature~\cite{Ratcliffe:1985mp}.
However, the above correlation functions Eqs.~(3,4) are special
projections of the general twist-three quark-gluon correlations,
and their evolutions are not directly available from the
already known results~\cite{Ratcliffe:1985mp}. Earlier attempts~\cite{Henneman:2001ev}
have been made to derive the evolution equations for the correlation
functions of Eqs.~(3,4), but were not complete.
On the other hand, from the large transverse momentum quark
Sivers function calculated in \cite{jqvy}, we would already obtain the evolution
equation for $T_F(x)$, since the collinear divergence in that calculation
will lead to the splitting function of $T_F(x)$. This
splitting function was confirmed by a complete calculation of next-to-leading
order QCD correction to the transverse-momentum weighted
spin asymmetry in Drell-Yan lepton pair production~\cite{yuan}. More comprehensive
evolution equations for $T_F(x)$, together with those for the three-gluon correlation
functions which are relevant to the single spin asymmetry observables
have recently been derived in~\cite{kang}. In this paper, we will extend these studies
to calculate the scale evolutions for the above four quark-gluon correlation functions.
Their contributions to the azimuthal angle distributions in Drell-Yan lepton
pair production in $pp$ collisions, and the relevant QCD factorization
analysis will be presented in a forthcoming publication~\cite{zhou}.

In our calculations, we will choose the light-cone gauge: $A^+=0$.
There are several advantages for this choice. First, the quark
Sivers function was previously calculated in the covariant
gauge~\cite{jqvy}. Our calculation in the light-cone gauge will
provide an important cross check for the results. Second, the
light-cone gauge is more convenient to calculate the evolution
equations for $\tilde g$ and $\tilde h$. In particular, the
evolution equations for $T_F(x)$ and $\tilde g(x)$ can be calculated
simultaneously. The only difference is that for $T_F(x)$ we have to
take a pole contribution for some diagrams, whereas for $\tilde g(x)$ we will not take
the pole (see the discussions below). Third, we can further choose a
particular boundary condition in the light-cone
gauge~\cite{Ji:2002aa}, which will greatly simplify the derivation.
We have also checked that the final results
do not depend on the boundary condition. According to the
quark distribution definition we have chosen above, it
is convenient to choose the retarded boundary condition, i.e.,
$A_\perp(-\infty^-)=0$. With this choice, the gauge link associated
with the TMD quark distributions in Eq.~(1) becomes unit, and their
contributions can be neglected~\cite{Ji:2002aa}. From this, we can
re-write the quark-gluon correlation functions $T_F(x)$ and $\tilde
g$ as
\begin{eqnarray}
&&T_F(x)=\int \frac{dy^-}{8 \pi^2 M} e^{ixP^+ y^-} \langle PS |
\bar{\psi}(0) n\!\!\!/ \epsilon^{\nu\mu}_\perp S_{\perp \nu}
i\partial_{\perp\mu} \psi_\alpha(y^-) | PS \rangle \ ,\\
&&\tilde g(x)=\int \frac{dy^-}{4 \pi M} e^{ixP^+ y^-} \langle PS |
\bar{\psi}_\beta(0) \gamma_5 n\!\!\!/  S_{\perp \mu}
i\partial_\perp^\mu \psi_\alpha(y^-) | PS \rangle\ ,
\end{eqnarray}
in the light-cone gauge with retarded boundary condition.
Similar expressions hold for other two correlation functions, $T_F^{(\sigma)}$ and
$\tilde h$. In the following calculations, we will focus on the derivation
for the evolution functions for $T_F$ and $\tilde g$, especially for $T_F$, and
those for $T_F^{(\sigma)}$ and $\tilde h$ can be obtained
accordingly.

To calculate the splitting function for the above two functions, we
have to take into account the contributions from the operators
$\left(\bar\psi\partial_\perp \psi\right)$ and $\left(\bar\psi A_\perp\psi\right)$~\cite{Ellis:1982cd},
because they are at the same order. Especially, because
of the contribution from $A_\perp$, the evolution of the above correlation
functions will involve more general twist-three functions: $G_D$ and $\tilde G_D$ or
$T_F(x_1,x_2)$ and $T_F^{(\sigma)}(x_1,x_2)$.
This is an important feature for the scale evolution
of the higher-twist distributions, such as that of the $g_T$ structure
function~\cite{Ratcliffe:1985mp}.

We plot the Feynman diagram contributions from the real gluon radiations
in Fig.~1, where (a) is the contribution
from the partial derivative on the quark field, and $(b-d)$ are those
from $A_\perp$ contributions. The virtual corrections only contribute
to partial derivative part, and they are easy to carry out.

\begin{figure}[t]
\begin{center}
\includegraphics[width=10cm]{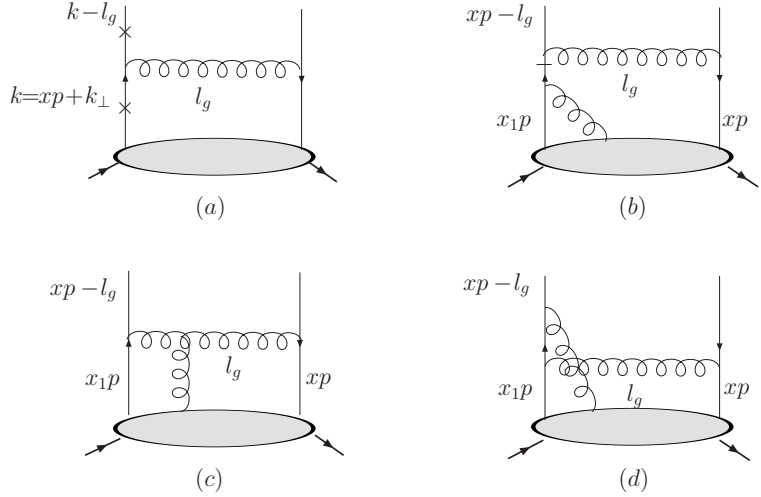}
\end{center}
\vskip -0.4cm \caption{Real gluon radiation contribution to
the evolution equations for the twist-three quark-gluon
correlation functions $T_F(x)$, $\tilde g(x)$, $T_F^{(\sigma)}(x)$, $\tilde h(x)$.} \label{fig1}
\end{figure}

We will perform the collinear expansion for the hard scattering part
to calculate the contribution from Fig.~1(a). The linear $k_\perp$ expansion
term combining with the quark field will lead to the quark-gluon correlation
function $T_F(x)$ and $\tilde g(x)$ in Eqs.~(9,10). In the collinear expansion
in terms of $k_\perp$, we
can fix the transverse momentum of the probing quark ($l_q$) or the radiated
gluon ($l_g$), because of momentum conservation and we are integrating over them
to obtain $T_F(x)$ and $\tilde g(x)$. We have also
checked that they will generate the same result. In the following
calculations, we choose $l_g$ being fixed in the collinear expansion.
This will avoid the collinear expansion of
the on-shell condition for the radiated gluon, and simplify the derivations.

For the $A_\perp$ contribution, we notice that $F^{+\mu}=\partial^+ A_\perp^\mu$ in
the light cone gauge. Therefore, one can relate the corresponding soft matrix to the correlation
function $T_F(x,x_1)$ in the following way,
\begin{eqnarray}
&& \frac{i}{x-x_1+i\epsilon}\int \frac{dy^-dy_1^-}{4 \pi} e^{ix_1P^+
y^-} e^{i(x-x_1)P^+y_1^-}\langle PS | \bar{\psi}_\beta(0^-)
n\!\!\!/ \epsilon^{\nu\mu}_\perp S_{\perp \nu} g{F^+}_\mu(y_1^-)
\psi_\alpha(y^-) | PS \rangle
\nonumber \\
&=&\int \frac{dy^-dy_1^-}{4 \pi }P^+ e^{ix_1P^+ y^-}
e^{i(x-x_1)P^+y_1^-}\langle PS | \bar{\psi}_\beta(0^-) n\!\!\!/
\epsilon^{\nu\mu}_\perp S_{\perp \nu} gA_{\perp\mu}(y_1^-)
\psi_\alpha(y^-) | PS \rangle \ .
\end{eqnarray}
In the above formula, the soft gluon pole appears in the first line
comes from the partial integration. The prescription of this
pole has been determined because we have chosen the retarded
boundary condition. For
the same reason, we have to regulate the light cone propagator in a
consistent manner, and the gluon propagator in Fig.~1(c) in the light cone gauge
with retarded boundary condition is given by~\cite{Ji:2002aa},
\begin{eqnarray}
D^{\alpha\beta}(l)=\frac{-i}{l^2+i\epsilon} \left ( g^{\alpha\beta} -
\frac{l^\alpha n^\beta +n^\alpha l^\beta}{l \cdot n + i\epsilon}
\right ) \ ,
\end{eqnarray}
where $l$ is the gluon propagator momentum entering the quark-gluon vertex in Fig.1(c).

Adding the contributions from the partial derivative and
$A_\perp$,  we reach our final formula for the $T_F(x)$ splitting
calculation,
\begin{eqnarray}
T_F^{(1)}(x_B)&=&\int dx d^2l_{g\perp} \frac{\partial}{\partial
k_\perp^\mu} \left \{ [ \hat{H}(k, l_g) p\!\!\!/ ] \times
l_{q\perp}^\mu \right \} |_{k_\perp=0} T_F(x,x)
\nonumber \\
&+&\int dx dx_1d^2l_{g\perp} \left \{ [ \hat{H}_\mu(xP, x_1P,l_g) p\!\!\!/
] \times l_{q\perp}^\mu \right \}\frac{1}{\pi} \frac{i}{x-x_1+i\epsilon}
T_F(x,x_1) \ ,
\end{eqnarray}
where the transverse spin vector has been integrated out, and the transverse index $\mu$
is not meant to be summed up.
$l_{q\perp}$ is the probing quark transverse momentum. In the above equation,
$\hat H(k,l_g)$ represents the hard partonic part in Fig.~1(a) with transverse
momentum dependence on $k_\perp$, and
$\hat H_\mu(xP,x_1P,l_g)$ the hard part for Figs.~1(b-d) with transverse
polarized gluon $A_{\perp\mu}$ insertion where all momenta are collinear.
We have to include both contributions to obtain a complete result.

We have similar expression for $\tilde g(x)$ splitting. The only difference is to
replace $\!\!\not p$ with $\gamma_5 \!\! \not p$ and similar replacement in
the hard parts in the above two terms. More over, because of simple Dirac
algebra, the first term is the same for both $T_F(x)$ and $\tilde g(x)$ which
comes from Fig.~1(a). Let us first discuss
this contribution. In the calculations, we have to perform the collinear expansion in terms of
$k_\perp^\mu$. Because of the momentum conservation, $l_q^\mu=k_\perp^\mu-l_g^\mu$,
we can separate the contribution from the explicit dependence on $k_\perp^\mu$,
\begin{eqnarray}
T_F^{(1)}|_{\rm Fig.1(a)}&=&\int dx T_F(x,x)d^2l_{g\perp}
\left\{\left[ \hat{H}(k, l_g) p\!\!\!/ \right]|_{k_\perp=0} -l_{g\perp}^\mu \frac{\partial}{\partial
k_\perp^\mu} \left[ \hat{H}(k, l_g) p\!\!\!/ \right]  |_{k_\perp=0}\right\} \ .
\end{eqnarray}
The first term is easy to derive, and its contribution will be
\begin{eqnarray}
\frac{\alpha_s}{2\pi} \int
\frac{ dx }{x} \frac{dl_{g\perp}^2}{l_{g\perp}^2} C_F \left(
\frac{1+z^2}{1-z} \right) T_F(x,x) \ .
\end{eqnarray}
where $z=x_B/x$, and the well-known splitting kernel appears.
This splitting kernel contains the end-point divergence,
which should be canceled out by the virtual diagram contributions.
After taking into account the virtual contribution, the end-point will be regulated by the
plus function,
\begin{eqnarray}
\frac{\alpha_s}{2\pi} \int
\frac{ dx }{x} \frac{dl_{g\perp}^2}{l_{g\perp}^2} C_F \left(
\frac{1+z^2}{(1-z)_+} +\frac{3}{2}\delta (1-z)\right) T_F(x,x) \ ,
\end{eqnarray}
where the plus function follows the definition of \cite{ap}.
To calculate the second term of Eq.~(14), we will do the collinear
expansion of the hard scattering part $\hat H(k,l_g)$ with $l_{g\perp}$
fixed. The transverse momentum $k_\perp$ flow can go through the
quark line in Fig.~1(a), for which we label as ``cross" in the diagram.
We can further simplify the derivation by using the following identity,
\begin{eqnarray}
\frac{\partial}{\partial k^\alpha}
\frac{i}{k\!\!\!/}=\frac{i}{k\!\!\!/} i \gamma^\alpha
\frac{i}{k\!\!\!/} \ ,
\end{eqnarray}
which essentially represents the application of the Ward identity. Applying
the above identity, we can relate the $k_\perp$ expansion in the
quark propagator and quark line to that with a transverse polarized
gluon insertion with zero momentum attachment. These contributions
shall be combined with those from Figs.~1(c-d). We will
discuss them below.

As we mentioned above, the contributions from Fig.~1(a) are the same
for the evolutions of $T_F(x)$ and $\tilde g(x)$. Therefore, the above
results apply for that of $\tilde g(x)$ too. However, the contributions
from Figs.~1(b-d) are different for $T_F(x)$ and $\tilde g(x)$. This
is because for this part, we have to take pole contribution to obtain the
splitting for $T_F(x)$, whereas for $\tilde g(x)$ we do not take pole contribution.
We first discuss their contributions to the evolution of $T_F(x)$
correlation function. Depending on the value of $x_g=x-x_1$
when we take the pole, these poles are called soft ($x_g=0$)
and hard ($x_g\neq 0$) poles, respectively.
The hard pole contribution only comes from the
light-cone propagator in Fig.~1(c), and its contribution is easy to
calculate. For this, we obtain
\begin{eqnarray}
T_F^{(1)}|_{\rm Fig.1(c)}^{\rm hp}&=&\frac{\alpha_s}{2\pi} \int
\frac{dx}{x} \frac{dl_{g\perp}^2}{l_{g\perp}^2} \frac{C_A}{2} \left(
\frac{1+z}{1-z } \right)T_F(xz,x) \ .
\end{eqnarray}
We emphasize that for the hard pole contribution the explicit factor
$1/(x-x_1)$ has been included in the above result, which is
finite because $x\neq x_1$.

On the other hand, the soft pole contribution comes from the explicit
pole in Eq.~(13) which leads to a delta function $\delta (x_1-x)$.
Because this pole results into zero gluon momentum insertion to the diagrams, we can combine
these contributions with the second term in Eq.~(14) as we mentioned above.
Therefore, we can add them together,
\begin{eqnarray}
-\int dx T_F(x,x)d^2l_{g\perp}l_{g\perp}^\mu
\left\{ \frac{\partial}{\partial k_\perp^\mu} \left[ \hat{H}(k, l_g) p\!\!\!/ \right]  |_{k_\perp=0}
-\left[\hat{H}_\mu(xP,xP, l_g) p\!\!\!/ \right] \right\} \ .
\end{eqnarray}
From this equation, we find that the contribution from Fig.~1(b) cancels
out that from the $k_\perp$ expansion on the quark line with momentum
``$k$", because they have the same color-factor but opposite signs.
The Fig.~1(d) and the $k_\perp$ expansion on the quark propagator
``$k-l_g$" are also the same but with different color-factor: color-factor
for Fig.~1(d) is $-1/2N_c$ whereas that for Fig.~1(a) is $C_F$. Their
total contributions will add up to a color-factor $C_A/2$. The same
color-factor $C_A/2$ appears for Fig.~1(c). Thus, the final result for
this contribution will be proportional to $C_A/2$. By applying the
identity of Eq.~(17) again, we can re-write this part of contribution
as
\begin{eqnarray}
&&-\frac{\alpha_s}{2\pi} \frac{C_A}{2}
\int dxT_F(x,x)  d^2l_{g\perp} \left [ \frac{ \partial}{\partial
l_{g\perp}^\mu } \hat H_0(xP,l_{g\perp} ) \right ] \times
(-l_{g\perp}^\mu)
\nonumber \\
&=&-\frac{\alpha_s}{2\pi}  \frac{C_A}{2} \int dxT_F(x,x)  d^2l_{g\perp}
\hat H_0(xP,l_{g\perp} )\nonumber\\
&=&-\frac{\alpha_s}{2\pi} \frac{C_A}{2}
\int dx \frac{dl_{g\perp}^2}{l_{g\perp}^2}\left(
\frac{1+z^2}{1-z} \right) T_F(x,x) \ ,
\end{eqnarray}
where $ \hat H_0$ represent the hard scattering part without color factor
and we have made use of the fact that the hard
part $\hat H_0(xP,l_{g\perp})\propto 1/l_{g\perp}^2$.
Again, the same splitting kernel appears.
Finally, there is also contribution from $\tilde T_F(x_1,x_2)$, which
only comes from the hard pole diagram Fig.~1(c).
Summing up all contributions, we obtain the scale evolution
equation for the diagonal part of the quark-gluon correlation function
$T_F(x_1,x_2)$,
\begin{eqnarray}
\frac{\partial}{\partial {\ln\mu^2}}
T_F(x_B,\mu^2)&=&\frac{\alpha_s}{2\pi} \int \frac{dx}{x} \left [
C_F \left \{ \frac{1+z^2}{(1-z)_+}+ \frac{3}{2} \delta(1-z) \right
\} T_F(x,x) \right .\
\nonumber \\
&&+ \left .\  \frac{C_A}{2} \left \{
\frac{1+z}{1-z } T_F(xz,x) -\frac{1+z^2}{1-z}  T_F(x,x)+ \tilde{T}_F(xz,x) \right \}
\right ] \ ,
\end{eqnarray}
which is consistent with that in Ref.~\cite{kang,yuan}.
The complete evolution equation for $T_F(x)$ shall also contain contributions
from the three-gluon correlation functions, which have been calculated in~\cite{kang}.

As we mentioned above, the contributions from Figs.~1(b-d) to the evolution
of $\tilde g(x)$ are different from that of $T_F(x)$. For $\tilde g(x)$ splitting,
we do not take pole contributions from these diagrams. For example,
we will not have cancelation between diagrams Fig.~1(b) and collinear
expansion of quark line ``$k$" of Fig.~1(a). More over, without taking
pole there will be an additional integral variable in the splitting function,
similar to that for the evolution of $g_T$ structure function~\cite{Ratcliffe:1985mp}.
The $A_\perp$ contribution from Figs.~1(b-d) can be transformed into $T_F$ and $\tilde T_F$, or
to $G_D$ and $\tilde G_D$~\cite{Jaffe:1991ra}. Because we do not take a pole
for the scattering amplitudes, the calculations for these diagrams are
straightforward. The partial derivative contribution from Fig.~1(a)
is similar to that for $T_F(x)$ calculation. This part depends on $\tilde g(x)$.
After adding all these contributions together, we  obtain the
evolution equation for $\tilde g(x)$,
\begin{eqnarray}
\frac{\partial \tilde{g}(x_B,\mu^2)}{\partial {\rm ln\mu^2}}
&=&\frac{\alpha_s}{2\pi} \int \frac{dxdy}{x }\left \{ \tilde{g}(x)\delta(y-x)
\left [ C_F  \left(\frac{1+z^2}{(1-z)_+}+ \frac{3}{2}\delta(1-z) \right )
 -\frac{C_A}{2}\frac{1+z^2}{1-z}\right ]\right.
\nonumber\\
&+&\ \!\!\!\tilde{G}_D(x,y)
\left [  C_F \left ( \frac{x_B^2}{x^2}+\frac{x_B}{y}-\frac{2x_B^2}{x y}-\frac{x_B}{x}-1 \right )
+ \frac{C_A}{2}
\frac{(x_B^2+x y)(2x_B-x-y)}{(x_B-y)(x-y)y} \right ]
\nonumber\\
&+& \left .\ \!\!\! G_D(x,y)
\left [ C_F \left (\frac{x_B^2}{x^2}+\frac{x_B}{y}-\frac{x_B}{x}-1 \right )
+\frac{C_A}{2}
\frac{x_B^2-xy}{(y-x_B)y} \right ]
\right \} \ ,
\end{eqnarray}
where again $z=x_B/x$, and the definitions of $G_D$ and $\tilde G_D$
follow that in~\cite{Jaffe:1991ra}.
The end-point singularity from $\tilde g(x)$ with color factor $C_A/2$ at right hand side of the
equation is canceled out by that from $\tilde G_D$ at the second line.
We further notice that we can replace $G_D$ and $\tilde G_D$ with
$T_F$ and $\tilde T_F$ at the right hand side by using the relations
between them~\cite{Ellis:1982cd,{Eguchi:2006mc}}. However,
we still have the $\tilde{g}(x)$ term at the right hand side of equation.
Although we can re-write $\tilde g(x)$ in terms of $\tilde G_D$ and
$\tilde T_F$~\cite{Eguchi:2006qz}, that will not eliminate its dependence
completely and the right hand side will depend on $\tilde G_D$, $\tilde T_F$
and $G_D$ instead.
Therefore, the evolution of $\tilde g(x)$ depends
on three functions: $\tilde g(x)$, $G_D(x,y)$ and $\tilde G_D(x,y)$. This feature is different
from that for $T_F(x)$, where it only depends on $T_F$ and $\tilde T_F$.
It may indicate the nontrivial QCD dynamics associated
with the evolution of the correlation function $\tilde g(x)$.
This has also been shown in its contribution to
the Drell-Yan dilepton azimuthal asymmetry in $pp$ scattering. We
leave this study in a future publication.

Since the derivation follows the similar procedure, we skip
the technique details and only list the final result for the evolution equation of
correlation functions $T_F^{(\sigma)}(x,x), \tilde{h}(x)$.
For $T_F^{(\sigma)}$, we have
\begin{eqnarray}
\frac{\partial}{\partial {\rm ln\mu^2}}
T_F^{(\sigma)}(x_B,\mu^2)&=&\frac{\alpha_s}{2\pi} \int
\frac{dx}{x} \left [ C_F \left \{ \frac{2z}{(1-z)_+}+ 2\delta(1-z)
\right \} T_F^{(\sigma)}(x,x) \right .\
\nonumber \\
&&\ \ \ \ \ \ \ \ \ \ + \left .\  \frac{C_A}{2} \left \{
\frac{2}{1-z } T_F^{(\sigma)}(xz,x) -\frac{2z}{1-z}
T_F^{(\sigma)}(x,x) \right \} \right ] \ ,
\end{eqnarray}
which is consistent with the large transverse
momentum Boer-Mulders function $h_1^\perp(x,k_\perp)$
calculated in~\cite{Zhou:2008fb}.
Accordingly, we obtain the evolution equation for $\tilde h$,
\begin{eqnarray}
\frac{\partial \tilde{h}(x_B,\mu^2)}{\partial {\rm ln\mu^2}}
&=&\frac{\alpha_s}{2\pi} \int \frac{dx dy}{x} \left\{\tilde{h}(x)\delta(y-x)\left[
C_F \left ( \frac{2z}{(1-z)_+}+ 2 \delta(1-z) \right )
-\frac{C_A}{2}\frac{2z}{1-z}\right ]\right. \nonumber\\
&&\left. +H_D(x,y)\left [ C_F \frac{2(x-y-x_B)}{y} +
\frac{C_A}{2}
\frac{2x_B(x_B x +x_By -x^2-y^2)}{(x_B-y)(x-y)y} \right ]
 \right \}
\end{eqnarray}
where the twist-three function $H_D(x,y)$ has been
introduced in the Ref.~\cite{Jaffe:1991ra}. Similar to
that of $\tilde g(x)$, the evolution of $\tilde h$ depends on $\tilde h$ and $H_D$.

In conclusion, we have derived the scale evolution for the transverse
momentum dependent quark-gluon correlation functions associated
with the four $k_\perp$-odd TMD quark distributions. We have
performed our calculations in light-cone gauge with a particular
boundary condition for the gauge potential, and we have checked
that our results do not depend on these boundary conditions. Our result on
the evolution of $T_F(x)$ confirms recent calculations~\cite{kang,yuan}.
The scale evolution for $\tilde g$ and $\tilde h$ reveals nontrivial QCD dynamics.
We hope this will stimulate further theoretical studies.

Meanwhile, we notice that the scale evolution for the general twist-three
operators have been calculated in the literature~\cite{Ratcliffe:1985mp}.
It will be interested to compare the evolution equations for the correlation
functions studied in this paper with these well-known results. Especially,
the evolution of the twist-three distribution $g_T(x)$ and its contribution to semi-inclusive
processes deserve further investigations. We will address these issues
in the forthcoming papers.

This work was supported in part by the U.S. Department of Energy
under contract DE-AC02-05CH11231 and the National Natural Science
Foundation of China under the approval No. 10525523. We are grateful
to RIKEN, Brookhaven National Laboratory and the U.S. Department of
Energy (contract number DE-AC02-98CH10886) for providing the
facilities essential for the completion of this work. J.Z. is
partially supported by China Scholarship Council.

\end {document}